\documentclass[sigconf,nonacm]{acmart}
\copyrightyear{2025}
\acmYear{2025}
\setcopyright{cc}
\setcctype{by}
\acmConference[SC '25]{The International Conference for High Performance Computing, Networking, Storage and Analysis}{November 16--21, 2025}{St Louis, MO, USA, LLNL-CONF-2010453}
\acmBooktitle{The International Conference for High Performance Computing, Networking, Storage and Analysis (SC '25), November 16--21, 2025, St Louis, MO, USA}
\acmDOI{10.1145/3712285.3759791}
\acmISBN{979-8-4007-1466-5/2025/11}

\usepackage{pervasives}

\usepackage{acmart-taps}

\usepackage{enumitem}

\usepackage[skip=0pt]{caption}

\begin{document}
\title{Bridging the Gap Between Binary and Source Based Package Management in Spack\\}

\author{John Gouwar}
\orcid{0000-0003-0494-7245}
\affiliation{\institution{Northeastern University}
\city{Boston}
\state{Massachusetts}
\country{USA}}
\email{gouwar.j@northeastern.edu}

\author{Gregory Becker}
\orcid{0000-0002-8472-4449}
\affiliation{\institution{Lawrence Livermore National Laboratory (LLNL)}
\city{Livermore}
\state{California}
\country{USA}}
\email{becker33@llnl.gov}

\author{Tamara Dahlgren}
\orcid{0009-0006-2279-9521}
\affiliation{\institution{Lawrence Livermore National Laboratory (LLNL)}
\city{Livermore}
\state{California}
\country{USA}}
\email{dahlgren1@llnl.gov}

\author{Nathan Hanford}
\orcid{0000-0002-2214-7447}
\affiliation{\institution{Lawrence Livermore National Laboratory (LLNL)}
\city{Livermore}
\country{USA}}
\email{nhanford@llnl.gov}

\author{Arjun Guha}
\orcid{0000-0002-7493-3271}
\affiliation{\institution{Northeastern University}
\city{Boston}
\state{Massachusetts}
\country{USA}}
\email{a.guha@northeastern.edu}

\author{Todd Gamblin}
\orcid{0000-0002-7857-2805}
\affiliation{\institution{Lawrence Livermore National Laboratory (LLNL)}
\city{Livermore}
\state{California}
\country{USA}}
\email{gamblin2@llnl.gov}

\renewcommand{\shortauthors}{Gouwar et. al}

\begin{abstract}
  Binary package managers install software quickly but they limit configurability
  due to rigid ABI requirements that ensure compatibility between binaries.
  Source package managers provide flexibility in building software, but compilation
  can be slow. For example, installing an HPC code with a new MPI implementation
  may result in a full rebuild. Spack, a widely deployed, HPC-focused package
  manager, can use source and pre-compiled binaries, but lacks a binary
  compatibility model, so it cannot mix binaries not built together.
  We present {\it splicing}, an extension to Spack that models binary compatibility
  between packages and allows seamless mixing of source and binary distributions.
  Splicing augments Spack's packaging language and dependency resolution engine to
  reuse compatible binaries but maintains the flexibility of source builds.
  It incurs minimal installation-time overhead and allows rapid
  installation from binaries, even for ABI-sensitive dependencies like MPI that
  would otherwise require many rebuilds.
\end{abstract}

\ccsdesc[500]{Software and its engineering~Software configuration management and version control systems}
\ccsdesc[300]{Software and its engineering~Interoperability}
\ccsdesc[100]{Software and its engineering~Domain specific languages}
\keywords{ABI-compatibility, Dependency management, Answer set programming,
  Binary reuse}

\maketitle


\section{Introduction}
\label{sec:introduction}

When developing and deploying high-performance software there is an inherent
trade off between building from source and using pre-compiled binaries. Building
from source provides full control over micro-architectural optimizations and
conditional features offered in bleeding-edge scientific software. However,
this customizability is often complex and time consuming, especially given the
large numbers of dependencies that must be compatibly built for HPC.
The more supported features and dependencies on third party libraries a package
has, the greater the difficulty in configuring and building the software
completely from source. The ability to use pre-compiled libraries can increase
the speed and ease of installation but at the cost of locking into potentially
limited configuration options that may not match the user's needs.

Spack \cite{gamblin2015spack} eliminates that sacrifice by providing
configurability with the relative ease of using pre-compiled libraries, either
those previously built from source by Spack or those that already exist on the
system. Spack-built libraries may be installed locally on the system or fetched
from an external binary cache. For example, package updates in the Spack repository
trigger creation of binaries that populate a public cache available to the Spack
community\footnote{https://spack.io/spack-binary-packages/}. Software already on
the system can include pre-compiled libraries that Spack cannot build, such as
vendor-specific MPI implementations, or hardware abstractions.

When using pre-compiled libraries that may have been built separately, it is
important to ensure that their \emph{application binary interfaces} (ABIs)
are compatible. ABI-compatible libraries agree on the compiled names of shared
symbols, the calling conventions for functions, and the size and layout of
shared types. Spack only allows {\it one} implementation of each dependency
in any installation's {\it directed acyclic graph} (DAG) of dependencies.
This trivially ensures ABI compatibility, because every compilation that
uses a given dependency will include the same headers, ensuring consistency.
An installation that built with a {\it different} version of the dependency
cannot occur in the same DAG. If the user requires a different version (or a
different implementation) of some dependency, all libraries built with the
prior version must be rebuilt.

On its face, this may not seem like too much of a restriction; however, it is
overly conservative and prevents reuse in cases where a different build of a
dependency would retain ABI-compatibility. To illustrate the consequences of
this restriction, consider deploying the widely used solver suite
\textsc{Trilinos} \cite{trilinos-website} on an HPE Cray
cluster. \textsc{Trilinos} has a dependency on an MPI implementation and HPE
Cray recommends using their specialized MPI implementation (\textsc{Cray MPICH})
on their hardware \cite{cray-docs}.  Notably, \textsc{Cray MPICH} is only
available on HPE Cray systems, but is ABI-compatible with the general purpose
\textsc{MPICH}.  Given Spack's current restrictions,
users would be forced to rebuild all of \textsc{Trilinos} on the cluster in
order to use \textsc{Cray MPICH} even though it is safe to build it in advance
against a compatible \textsc{MPICH} and simply link against
\textsc{Cray MPICH}. This becomes especially painful when trying to {\it distribute}
MPI binaries. Ideally, we would build a stack of software with the publicly
available \textsc{MPICH} and allow it to be installed without rebuilding on
on any system with an ABI-compatible, optimized MPI implementation. Because
Spack currently cannot describe when two packages are ABI-compatible,
and because it cannot represent heterogeneous binaries built with {\it different}
dependency implementations, it cannot effectively reuse binaries that depend
on MPI, and it must rebuild.

In this paper, we present an extension to Spack called \emph{splicing}.
We augment Spack with a model of package ABI-compatibility and a model of
heterogeneous dependencies that may not have been initially compiled together.
Splicing allows Spack to replace dependencies of pre-compiled libraries
with ABI-compatible substitutes. We add a single directive,
\texttt{can\_splice}, to Spack's domain specific language (DSL), which allows
developers to specify when compiled configurations of libraries are ABI-compatible.
We also extend Spack's dependency resolution mechanism to synthesize
ABI-compatible splices when using pre-compiled libraries. These features create
an additional dependency structure on the Spack ecosystem that allows for
specifying the ABI compatibility of a package's compiled representations and how
those can be safely replaced as the dependency of another pre-compiled
library.
In summary, we provide the following contributions:
\begin{enumerate}
\item A formal model of \textit{splicing}, a representation of binary
  re-linking in Spack that maintains full build provenance;

\item An augmentation to Spack's packaging language for specifying ABI
  compatibility between package configurations;

\item An implementation of automatic splice synthesis in Spack's
  solver and re-linking in Spack's installer; and

\item An empirical evaluation of scaling, performance, and correctness of
  dependency resolution augmented with splicing.
\end{enumerate}



\section{Background}
\subsection{API vs. ABI Compatibility}
There are two forms of application interface compatibility: program and binary.
An application programming interface (API) is a source-level contract describing
the interface that a library provides to others. It includes the signatures of
functions and names of exported types. A package X is API-compatible with a
package Y if X implements a superset of Y's interface. That is, any package
which uses Y's interface may safely use X's implementation of that interface,
even though X may provide additional functionality. For C libraries, this
information is encapsulated in header files, and API-compatible packages
implement equivalent headers. An
application binary interface (ABI) describes the boundary between two
independently compiled software artifacts. This includes the compiler-mangled
names of exported symbols and the layout of exported types. A compiled package X
is ABI-compatible with a compiled package Y if it is API compatible {\it and}
its compiled representation has the same mangled names and type layouts.

While API compatibility is necessary for ABI compatibility, it is not
sufficient. For example, if a user-defined type in a particular interface is
opaque, like MPI's \texttt{MPI\_Comm}, code that treats this type opaquely
at the source level may be safely compiled against any API-compatible
implementation of MPI, but once compiled it can only be safely used with code that
implements the type in the \emph{same} way that it was compiled. The C ABI for
language-specified types (\textit{e.g.,} ints, floats, structs \textit{etc.}) is rigorously
specified for any particular architecture. Opaque type incompatibility at the
binary level actually arises between \textsc{Open MPI}~\cite{gabriel2004open}
and \textsc{MPICH}~\cite{gropp1996mpich}.
While \textsc{Open MPI} implements \texttt{MPI\_Comm} as an incomplete
struct pointer, \textsc{MPICH} implements it as a 32-bit integer~\cite{hammond2023mpi}.
Binaries compiled against one implementation cannot safely use the other.

\subsection{Package Managers}
\label{sec:background-pm}
Package managers provide tools to mitigate the complexity of installing
software. They can solve user-imposed constraints, handle dependencies,
build packages from source, and install precompiled binaries.
Most package managers have either a \emph{binary} or \emph{source} based
deployment paradigm. Binary package managers, like those included with
most major Linux distributions (Gentoo Linux being a notable exception), select
and download software versions as pre-compiled binary artifacts (\textit{e.g.,} shared
libraries, executables, \textit{etc.}). Binary package managers allow for quick
installation, but they limit package versions to specific pre-compiled
configurations. They also require that all expressed dependency constraints be between
ABI-compatible versions, since the packages installed by a user will be compiled
independently.

Source-based package managers download source code, build it, and deploy the
resulting software. In addition to installing files, they must manage the
associated build or interpretation. Generally, all dependent packages of a
source build will be compiled consistently, i.e., only a single version of any
dependency will be used, and flags are kept consistent across all
builds. Package developers can thus express dependencies at the API level.  ABI
compatibility follows by virtue of the consistent compilation model.  While the
lack of ABI modeling allows for more expressiveness, updates to packages with
many dependents trigger long cascades of recompilation. Updating to an ABI
incompatible version of a package in an existing dependency tree will require
all \emph{dependent packages} to rebuild against the new version. These rebuilds
in the dependent packages can then require rebuilds of their own dependent
packages, with required rebuilds potentially cascading to the whole dependency
tree. This need to \emph{rebuild the world} can have a chillling effect on a
developer's willingness to update a single dependency.  Furthermore, not all
source code is open and developers may have to rely on some software that is
only available as a binary, especially in the HPC space. This means that source
based package managers will also build with binary packages out of necessity,
even if the source code was never available to the package manager.



\section{Spack}
\label{sec:spack}

Spack is a source {\it and} binary package manager, widely used in HPC. This
section describes Spack's \textit{packages}, \textit{specs},
\textit{concretizer}, and \textit{binary relocation}, core components of
configuration and dependency resolution that provide needed context for
splicing.

\subsection{Specs}
\label{sec:spack-specs}
\textit{Specs} are the core of Spack; they allow users to concisely specify
build configuartions and constraints. Each spec has six attributes:
\begin{enumerate*}
  \item the package name;
  \item the version to build;
  \item variant values (\textit{i.e.,} compile-time options);
  \item the target operating system;
  \item the target microarchitecture;
  \item and the specs of the packages it depends on (and their attributes).
\end{enumerate*}

\aptLtoX[graphic=no,type=html]{\begin{table}
  \begin{tabular}{|c|c|c|}
    \hline
    Spec Sigil & Example & Meaning \\
    \hline
    @ & \texttt{hdf5@1.14.5} & Require version \\
    \hline
    + & \texttt{hdf5+cxx} & Require variant \\
    \textasciitilde & \texttt{hdf5\textasciitilde{}mpi} & Disable variant \\
    \hline
    \textasciicircum & \texttt{hdf5 \textasciicircum{}zlib} & Depends on (link-run)\\
    \% & \texttt{hdf5\%clang} & Depends on (build)\\
    \hline
    \texttt{key=value} & \texttt{hdf5 target=icelake} \newline
                       \texttt{hdf5 api=default}  & Require variant or target
                                                  to hold value \\
    \hline
  \end{tabular}
  \caption{Examples of spec syntax, and their meaning.}
  \Description{A table detailing the spec CLI syntax used by Spack.}
  \label{fig:spec-syntax}
\end{table}}{\begin{table}
  \begin{tabularx}{1.0\columnwidth}{|c|X|X|}
    \hline
    Spec Sigil & Example & Meaning \\
    \hline
    \texttt{@} & \texttt{hdf5@1.14.5} & Require version \\
    \hline
    \texttt{+} & \texttt{hdf5+cxx} & Require variant \\
    \texttt{\textasciitilde} & \texttt{hdf5\textasciitilde{}mpi} & Disable variant \\
    \hline
    \texttt{\textasciicircum} & \texttt{hdf5 \textasciicircum{}zlib} & Depends on (link-run)\\
    \texttt{\%} & \texttt{hdf5\%clang} & Depends on (build)\\
    \hline
    \texttt{key=value} & \texttt{hdf5 target=icelake} \newline
                       \texttt{hdf5 api=default}  & Require variant or target
                                                  to hold value \\
    \hline
  \end{tabularx}
  \caption{Examples of spec syntax, and their meaning.}
  \Description{A table detailing the spec CLI syntax used by Spack.}
  \label{fig:spec-syntax}
\end{table}}


%
Spack provides a concise syntax for specs, and Table~\ref{fig:spec-syntax}
shows a subset of simple examples needed to understand this paper.
A user typically provides an incomplete description,
or \textit{abstract spec}, by not explicitly constraining all
attributes. In contrast, a spec with all attributes set is called a
\textit{concrete spec}. Concrete specs can be built and installed, as they
contain all information the build environment might need to query in order
to configure and build.
The rest of the paper is concerned mainly with concrete specs.

Specs are recursive in that dependencies (attribute (6) above) are specs
themselves. They are therefore represented in memory as directed acyclic
multigraphs. Nodes correspond to packages with their attributes, a directed
edge $(A, B)$ denoting that package $A$ \textit{depends on} package $B$, and two
edge sets corresponding to build, and link-run dependencies.  Taking the union
of all edge sets and forming a single DAG admits a performant hashing scheme on
specs, allowing for efficient reasoning about equality between specs.

\subsection{Packages}
\label{sec:spack-packages}
A Spack \textit{package} defines the build process for a software
product using available options. It defines a combinatorial space of build
options that can include software versions, optional features,
compile-time flags, dependencies on other packages, potential conflicts with
other packages, and environmental constraints (\textit{e.g.,} requiring an x86\_64
system, requiring a CUDA installation, \textit{etc.}).
The software is described in a DSL embedded in Python that uses a Python
class to represent every build configuration, a \textit{parameterized} build
process, and the artifacts (\textit{e.g.,} tarballs and patches) needed to build.
%


\begin{figure}

\input{code10.tex}
\caption{An example \texttt{package.py} file demonstrating constraints in
    Spack's embedded DSL. This package also contains our new augmentation to the
    packaging language, \texttt{can\_splice}, which we introduce in
    section~\ref{sec:impl-new-directive}.}
  \label{fig:example-pkg}
  \Description{A snippet of Python code demonstrating the Spack packaging DSL.}
\end{figure}


%
Figure~\ref{fig:example-pkg} shows a simple Spack package. Package
configuration features, such as \texttt{version}, \texttt{depends\_on},
and \texttt{variant}, are specified with \textit{directives}. Directives
serve to define the configuration space of a package and may introduce
constraints on the package or its dependencies.\footnote{Some directives
  simply attach metadata, such as the Github usernames of the maintainers,
  licensing information, or websites associated with the package.}
In contrast to most package managers, where a package description specifies
a single software configuration, Spack packages are {\it conditional}, and most
directives accept a {\tt when} argument. For example, the \texttt{depends\_on}
directive on line 13 constrains the {\tt zlib} dependency to version {\tt 1.2}
{\it when} the example package is at version {\tt 1.0.0}. Similarly, the
directive on line 15 says that
{\tt example} version {\tt 1.1.0} requires a newer {\tt zlib} version: {\tt 1.3}.
Constraints in package directives are specified using the spec syntax
(Section~\ref{sec:spack-specs}). Directives below \texttt{\#\#\#}
are part of our DSL extension (section~\ref{sec:impl-new-directive}).

For our purposes, there are two major classes of dependencies in Spack: build
and link-run. Build dependencies are packages that a node needs to
execute its build process; these can be compilers (\textit{e.g.,} \textsc{GCC}, or
\textsc{clang}), (meta-)build systems(\textit{e.g.,} \textsc{autotools},
\textsc{CMake}, or \textsc{Ninja}), or even interpreters for build-time glue
code (\textit{e.g.,} \textsc{Python} or \textsc{Perl}). Link-run dependencies are the
packages that a node needs either at compile-time for linking (\textit{i.e.,} shared
object files for dynamic linking and assembly code for static linking), or at
runtime to use either as a subprocess or as a dynamically loaded library.
\subsection{The Concretizer}
\label{sec:spack-concretizer}
Spack's dependency resolver, or {\it concretizer}, takes abstract specs
requested by a user and produces concrete specs that are valid according
to constraints in package files. Since specs are recursive, this also
includes resolving concrete specs for all of a package's
dependencies. Consider the following abstract spec for the package from
fig.~\ref{fig:example-pkg}, \texttt{example@1.0.0}; one such concretization could be:

\begin{footnotesize}
\par\vspace{3em}\noindent


\begin{Verbatim}[commandchars=\\\{\},fontsize=\footnotesize,frame=single]
example@1.0.0 +bzip arch=linux\PYZhy{}centos8\PYZhy{}skylake
    \PYZca{}bzip2@1.0.8 \PYZti{}debug+pic+shared arch=linux\PYZhy{}centos8\PYZhy{}skylake
    \PYZca{}zlib@1.2.11 +optimize+pic+shared arch=linux\PYZhy{}centos8\PYZhy{}skylake
    \PYZca{}mpich@3.1 pmi=pmix arch=linux\PYZhy{}centos8\PYZhy{}skylake
\end{Verbatim}


\vspace{0.3em}
\end{footnotesize}

Dependency resolution when considering only compatible versions is known to be
NP-complete~\cite{dicosmo2006edos,cox2016version}, and Spack must also decide
compatible build options, operating systems, and micro-architectures.  Spack
attempts to maximize its reuse of already-built components, either
previously installed locally or present in some remote cache, which we
collectively call \textit{reusable specs}. To ensure completeness and optimality
(\textit{i.e.,} always finding the optimal solution if one exists) while maintaining
tractability, Spack implements its concretizer using the Answer Set Programming
(ASP) system Clingo~\cite{gebser2016potassco,gamblin2022asp}. ASP is a logic
programming paradigm, similar to Datalog, where problems are specified as
first-order logic programs (\textit{i.e.,} facts and deductive rules with variables)
extended with a non-deterministic choice construct. ASP is distinguished from
other logic programming paradigms by its lack of an operational semantics;
instead, ASP programs are first \textit{grounded} into propositional logic
programs without variables, and then solved for their \textit{stable models}
using techniques from the SAT/SMT solving community such as the classic
Davis–Putnam–Logemann–Loveland (DPLL) algorithm along with modern extensions
like Conflict-Driven Clause Learning (CDCL)~\cite{davis1960computing,davis1962machine,moskewicz2001chaff}. Since a given program may have many
models in the presence of non-deterministic choices, Clingo allows
optimization objectives to be defined over models of a program.

Spack's concretizer is implemented in three stages. First, constraints from the
package classes, reusable specs, and the user-provided abstract specs are
compiled to an encoding of facts in ASP. Then, these facts are included in a
logic program describing the formal constraints of spec concretization (\textit{e.g.,}
every spec must be resolved to exactly one version, every variant must have a
chosen value, \textit{etc.}) and optimization objectives (\textit{e.g.,} maximize the reusable
specs, use the newest satisfying version of every package, \textit{etc.}), which is
then run with Clingo to solve for stable models. Finally, the optimal model
is processed by an interpreter to construct concrete specs for further use by
Spack. Notably, this output includes which specs are reused (\textit{i.e.,} those which have
already been built), and which ones must be built from source.
Further details on the concretizer implementation, its optimization criteria,
and the implementation of package reuse can be found in prior work~\cite{gamblin2022asp}.

\subsection{Binary Relocation}
\label{sec:spack-relocation}
Once Spack builds a package, it needs to ensure that each executable and library knows
where to find its dependencies. On systems where there is only one version of each
library, managed by the system package manager, the search path for
libraries is usually stored either in global system configuration (\textit{e.g.,} {\tt ldconfig} or {\tt
  ld.so.conf}), or in an environment variable (\textit{e.g.,} \texttt{LD\_LIBRARY\_PATH}). A
naive install algorithm would simply append installation paths for Spack-built binaries
to system search paths. This could lead to unintended crashes in both system
libraries and Spack-built libraries, since each could attempt to access unexpected,
incompatible libraries.
Taking inspiration from the Nix package
manager~\cite{dolstra2004nix,dolstra2008nixos},
Spack instead uses RPATHs~\cite{gamblin2015spack}.
RPATHs allow for embedding the location of shared libraries directly into a
binary without environment modification. All
Spack packages are installed in a user-defined prefix and all dependencies are
embedded as RPATHs.

Unlike Nix, Spack supports {\it binary relocation}. Spack build caches can be installed
in any location, including user home directories. The deployment locations of a binary
package and its dependencies may {\it not} be the same at runtime as at build time. To
ensure correct execution, Spack rewrites references to installation locations in
installed files. It creates a mapping from binaries' original
install locations to new locations, and it patches all occurrences of the old locations.
For simple cases, where the install location is shorter than the build location, Spack
uses simple patching logic. For more complex installs, it can use the \texttt{patchelf}
tool to lengthen paths~\cite{dolstra-patchelf}.
Spack can build binaries on a node's local filesystem,
relocate them to a network file system, push them to a remote server for hosting
reusable binaries, and install them again on a separate cluster.



\section{Splicing}


\begin{figure}
  \includegraphics{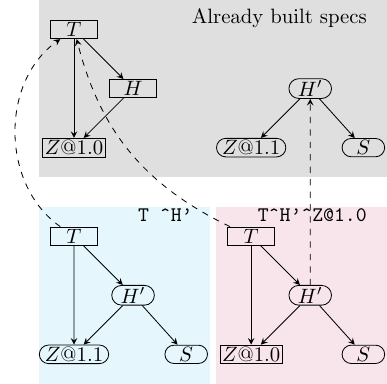}
  \caption{An example of how splicing combines spec-DAGs while maintaining build
    provenance. The specs in gray, \texttt{T \textasciicircum{}H \textasciicircum{}Z@1.0} (rectangular nodes) and
    \texttt{H' \textasciicircum{}S \textasciicircum{}Z@1.1} (rounded nodes), are already built. A spec
    conforming to \texttt{T \textasciicircum{}H'} can be satisfied with a
    splice of the already built \texttt{H'} into already built \texttt{T} (blue
    background). A spec conforming to \texttt{T \textasciicircum{}H' \textasciicircum{}Z@1.0} can be satisfied
    by splicing \texttt{Z@1.0} back into the previous spec (red background). The
  build provenance for spliced specs are denoted with dashed lines.}
\label{fig:splice-example}
\Description{A graphical representation of transitive and intransitive splices.}
\end{figure}


Supporting ABI compatibility presents two key package management opportunities.
The first is to avoid the unnecessary cascade of rebuilds for transitive
dependent packages when swapping ABI-compatible dependencies. Similarly,
deploying a package binary built with a reference library (like \textsc{MPICH})
onto systems with custom ABI-compatible implementations should not trigger rebuilds.

As discussed in section~\ref{sec:background-pm}, updating a dependency in a source
build can cause a series of package rebuilds for everything that directly
and transitively depends on it even when the updated package is ABI-compatible
with the version of the package that it is updating.  Instead, only the updated
package needs to be rebuilt and the RPATHs of all dependents updated to the
new binary.

The deployment scenario is even simpler.
Recall from section~\ref{sec:introduction} the problem of deploying a package on a
cluster using \textsc{Cray MPICH} when the package binary was built against
a general implementation {MPICH} on a build server.
Notably, \textsc{Cray MPICH} is usually a binary that
exists only on the cluster.
Assuming the general implementation conforms to the \textsc{Cray MPICH} ABI,
deployment of the package binary should only require that it point to \textsc{Cray MPICH}.

These scenarios introduce two challenges for Spack:
\begin{enumerate}
\item How should Spack represent binaries built against one
  implementation of an interface and deployed against another?
\item How should developers specify ABI-compatibility information to Spack such
  that these solutions are automated?
\end{enumerate}
In this section we present \textit{splicing} as a solution to the first
challenge and focus on its mechanics and representation. We also present a
solution to the second challenge of automating splicing in
section~\ref{sec:implementation}.  Splicing is a model of describing package
installations built against one shared library and deployed against
another, while maintaining full build provenance.

\subsection{The Mechanics of Splicing}
\label{sec:splice-mechanics}
We present the mechanics of splicing through a stripped down synthetic scenario,
as depicted in fig.~\ref{fig:splice-example}: There are two pre-compiled packages,
\texttt{T} and \texttt{H'}, conforming to \texttt{T \textasciicircum{}H \textasciicircum{}Z@1.0} (rectangular nodes)
and \texttt{H' \textasciicircum{}S \textasciicircum{}Z@1.1} (rounded nodes). The user is requesting a package
installation conforming to \texttt{T \textasciicircum{}H'}, and Spack knows that \texttt{H} and
\texttt{Z@1.0} are ABI-compatible with \texttt{H'} and \texttt{Z@1.1},
respectively. Currently, the best Spack could do to satisfy the request would be
to recompile \texttt{T} against \texttt{H'}; however, our goal is to satisfy the
user's request without recompiling.

At the level of spec DAGs, the solution is as simple as replacing the \texttt{T}
dependency on \texttt{H} with the full spec for \texttt{H'}, which we call splicing;
however, there are a few subtleties to acknowledge. The first is the
handling of the shared \texttt{Z} dependency. Simply copying in the full spec for
\texttt{H'} would leave two copies of the \texttt{Z} dependency at different
versions.  Since there can only be one version of any dependency in the link-run
graph for a spec, splicing must have a way to ``break ties.'' In general, we
consider splices to be \emph{transitive}, namely that splicing a dependent spec
also replaces all shared dependencies between the root and the spec being
spliced in.

If the user instead requested \texttt{T \textasciicircum{}H' \textasciicircum{}Z@1.0} (\textit{i.e.,}
explicitly constraining \texttt{Z}), this can be solved with another
splice. Splicing \texttt{Z} back into the the initial transitively spliced spec
(blue background in fig.~\ref{fig:splice-example}), would produce an
\emph{intransitive} splice (red background in fig.~\ref{fig:splice-example}), where
\texttt{Z@1.1} dependency for \texttt{H'} is replaced with \texttt{Z@1.0}, and the
dependency for \texttt{T} is restored.

The second subtlety is related to the \emph{build provenance} of spliced
specs. Note that \texttt{T \textasciicircum{}H' \textasciicircum{}Z@1.1} \emph{could} have been how the related
binaries were built, and thus a naive implementation attempting to reproduce
this spec would build it directly. Yet, this is not actually how the binaries
corresponding to this spec were built. First, \texttt{T \textasciicircum{}H \textasciicircum{}Z@1.0} and
\texttt{H' \textasciicircum{}Z@1.1} were built, and then they were spliced together. In order to
maintain reproducibility, Spack needs to distinguish spliced specs
from non-spliced specs and should be able reproduce the whole process of how a
spliced spec was constructed by building the initial specs and then splicing
them together, rather than just building the resulting spec. Therefore, we
augment spliced specs with a reference to the spec from which they are derived
that we call the \emph{build spec} (denoted by the dashed line in fig.~\ref{fig:splice-example}).

The final subtlety is the handling of build dependencies (not
pictured). Splicing is only a consideration for concrete specs which have
already been built (\textit{i.e.,} there is an existing binary to be
relinked). On the other hand, recall that build dependencies need not be present
\emph{after} being used to build the spec's binary.
Therefore, build dependencies are removed from the spec whose dependency
has been spliced, since they are no longer relevant to the runtime
representation. By keeping around the aforementioned build spec, removing these
build dependencies from the spliced spec does not sacrifice reproducibility.

\subsection{Patching Spliced Binaries}
\label{sec:splice-relocation}
Recall from section~\ref{sec:spack-relocation} that all Spack requires for binary
relocation is a mapping from the current installation paths of shared libraries
to the desired relocation paths of those libraries. This machinery can easily be
reused for patching binaries from spliced specs by generalizing
\emph{relocation} to \emph{rewiring}. Instead of changing the paths of the same
shared library to different locations, the shared library of the old node in the
spec before it was spliced is patched to be the new shared library of the
analogous node after it was spliced. This rewiring demonstrates the need for the
build spec in a spliced spec, since it allows for Spack to determine the mapping
between how the binary was initially installed to how it needs to be patched in
order to conform to the spliced spec. Build specs also make it simple to
determine whether a spec has been spliced at all, since only spliced specs will
have an associated build spec.



\section{Automatic Splicing}
\label{sec:implementation}
In this section we describe the implementation of automatic splicing in Spack.
The concretizer synthesizes spliced solutions from pre-installed specs when
package developer-specified declarative constraints on splice validity are
satisfied. As described in previous sections, this allows specs from a
buildcache to be built using general implementations of common interfaces, like
\textsc{MPICH}, and deployed without user intervention against specialized
implementations, such as \textsc{CrayMPICH}.  It also enables updating
dependencies in a way that does not require \emph{rebuilding the world}.

Several changes needed to be made to Spack. A new directive was added
to the packaging language for expressing
ABI-compatibility (section~\ref{sec:impl-new-directive}). The ASP encoding of
pre-installed specs needed to be changed to allow for splicing (section~\ref{sec:impl-new-encoding}).
Finally, logic was added to the concretizer to synthesize and execute splices
(section~\ref{sec:impl-splice-concretizer}). Section~\ref{sec:impl-basic-asp} provides an
introduction to the ASP encoding and techniques used in Spack's
concretizer.\footnote{Most ASP examples are simplified to the core logic
  necessary to understand the approach in this work. Spack's full concretizer
  has a number of indirections and augmentations to support a diverse array of
  features that are orthogonal to splicing and binary reuse.}

\subsection{ASP in the concretizer}
\label{sec:impl-basic-asp}
Recall from section~\ref{sec:spack-concretizer} that the ASP portion of Spack's
concretizer takes four inputs: compiled ASP facts of the user's requested spec,
compiled ASP facts from all package definition classes, an encoding of the
concrete specs that it can reuse, and a logic program implementing
Spack's concretization semantics. Specs are encoded through two relations:
\texttt{node} and \texttt{attr}. The unary \texttt{node} relation simply takes the name
of the package that it is encoding to indicate its presence. The higher-order,
variable arity \texttt{attr} relation describes how various attributes are
attached to nodes. The first argument of \texttt{attr}
is always the name of the attribute, the second is the node to which it applies,
and all other arguments depend on the particular attribute. By using a
higher-order \texttt{attr} relation, rather than individual relations for each
potential attribute, common patterns that relate to all of the potential
attributes can be reused (such as \texttt{impose} later in this section). The
following is the encoding of \texttt{example+bzip \textasciicircum{}bzip2}:
\par\vspace{0.5em}\noindent


\begin{Verbatim}[commandchars=\\\{\},fontsize=\footnotesize,frame=single]
\PY{n+nf}{node}\PY{p}{(}\PY{l+s+s2}{\PYZdq{}example\PYZdq{}}\PY{p}{)}\PY{p}{.}
\PY{n+nf}{node}\PY{p}{(}\PY{l+s+s2}{\PYZdq{}bzip2\PYZdq{}}\PY{p}{)}\PY{p}{.}
\PY{n+nf}{attr}\PY{p}{(}\PY{l+s+s2}{\PYZdq{}depends\PYZus{}on\PYZdq{}}\PY{p}{,}\PY{n+nf}{node}\PY{p}{(}\PY{l+s+s2}{\PYZdq{}example\PYZdq{}}\PY{p}{)}\PY{p}{,}\PY{n+nf}{node}\PY{p}{(}\PY{l+s+s2}{\PYZdq{}bzip2\PYZdq{}}\PY{p}{)}\PY{p}{,}\PY{l+s+s2}{\PYZdq{}link\PYZhy{}run\PYZdq{}}\PY{p}{)}\PY{p}{.}
\PY{n+nf}{attr}\PY{p}{(}\PY{l+s+s2}{\PYZdq{}variant\PYZdq{}}\PY{p}{,}\PY{n+nf}{node}\PY{p}{(}\PY{l+s+s2}{\PYZdq{}example\PYZdq{}}\PY{p}{)}\PY{p}{,}\PY{l+s+s2}{\PYZdq{}bzip\PYZdq{}}\PY{p}{,}\PY{l+s+s2}{\PYZdq{}True\PYZdq{}}\PY{p}{)}\PY{p}{.}
\end{Verbatim}


\vspace{0.5em}
Notably, the spec encoded above is not yet concrete, since not all attributes
have been set (\textit{e.g.,} the versions of the packages). The logic program
implementing the concretizer describes how to take user provided abstract
specs and the compiled package definitions to set \emph{all} attributes on the
nodes. Attributes range not only over the six for concrete specs
described in section~\ref{sec:spack-specs}, but also can describe various integrity
properties of the whole spec DAG as it relates to a particular node.
There are over twenty different attributes considered in the concretizer in all.

The directives from package definitions are compiled to analogous ASP facts
describing constraints on the solution space. The following is an example of the
facts related to versions for the package from fig.~\ref{fig:example-pkg}, declaring
two unconditional versions for the package \texttt{1.1.0} and \texttt{1.0.0}:
\par\vspace{0.5em}\noindent


\begin{Verbatim}[commandchars=\\\{\},fontsize=\footnotesize,frame=single]
\PY{n+nf}{pkg\PYZus{}fact}\PY{p}{(}\PY{l+s+s2}{\PYZdq{}example\PYZdq{}}\PY{p}{,}\PY{n+nf}{version\PYZus{}declared}\PY{p}{(}\PY{l+s+s2}{\PYZdq{}1.1.0\PYZdq{}}\PY{p}{)}\PY{p}{)}\PY{p}{.}
\PY{n+nf}{pkg\PYZus{}fact}\PY{p}{(}\PY{l+s+s2}{\PYZdq{}example\PYZdq{}}\PY{p}{,}\PY{n+nf}{version\PYZus{}declared}\PY{p}{(}\PY{l+s+s2}{\PYZdq{}1.0.0\PYZdq{}}\PY{p}{)}\PY{p}{)}\PY{p}{.}
\end{Verbatim}


The logic program then has the related rule for selecting versions which can
be read as ``choose exactly one version for each node from the
declared versions.'':

\begin{Verbatim}[commandchars=\\\{\},fontsize=\footnotesize,frame=single]
\PY{l+m+mi}{1} \PY{p}{\PYZob{}} \PY{n+nf}{attr}\PY{p}{(}\PY{l+s+s2}{\PYZdq{}version\PYZdq{}}\PY{p}{,}\PY{n+nf}{node}\PY{p}{(}\PY{n+nv}{Name}\PY{p}{)}\PY{p}{,}\PY{n+nv}{Ver}\PY{p}{)}
    \PY{l+s+sAtom}{:} \PY{n+nf}{pkg\PYZus{}fact}\PY{p}{(}\PY{n+nv}{Name}\PY{p}{,} \PY{n+nf}{version\PYZus{}declared}\PY{p}{(}\PY{n+nv}{Ver}\PY{p}{)}
 \PY{p}{\PYZcb{}} \PY{l+m+mi}{1} \PY{p}{:\PYZhy{}}  
  \PY{n+nf}{attr}\PY{p}{(}\PY{l+s+s2}{\PYZdq{}node\PYZdq{}}\PY{p}{,}\PY{n+nf}{node}\PY{p}{(}\PY{n+nv}{Name}\PY{p}{)}\PY{p}{)}\PY{p}{.}
\end{Verbatim}


\vspace{0.5em}
There are analogous corresponding package facts and concretizer rules for each
associated attribute for concrete spec (\textit{e.g.,} variant values, dependencies, \textit{etc.}).

\subsubsection{Handling conditional constraints}
\label{sec:concretizer-imposition}
Many package directives have a \texttt{when} argument for expressing
conditional constraints. These require a little more machinery to express in
ASP. Each condition is given a unique identifier, the requirements
for the condition to be satisfied, and the constraint imposed on the resulting
models if those conditions hold. Below is the compiled representation of the
conditional dependency from line 11 of the package in fig.~\ref{fig:example-pkg}. It
declares a condition with unique identifier \texttt{"x153"}, with conditions that
there is a node \texttt{"example"} with variant \texttt{"bzip"} set
to \texttt{"True"}. The condition then imposes the constraint that the \texttt{"example"}
node also depends on \texttt{"bzip2"}.

\begin{Verbatim}[commandchars=\\\{\},fontsize=\footnotesize,frame=single]
\PY{n+nf}{condition}\PY{p}{(}\PY{l+s+s2}{\PYZdq{}x153\PYZdq{}}\PY{p}{)}\PY{p}{.}
\PY{n+nf}{condition\PYZus{}requirement}\PY{p}{(}\PY{l+s+s2}{\PYZdq{}x153\PYZdq{}}\PY{p}{,}\PY{l+s+s2}{\PYZdq{}node\PYZdq{}}\PY{p}{,}\PY{l+s+s2}{\PYZdq{}example\PYZdq{}}\PY{p}{)}\PY{p}{.}
\PY{n+nf}{condition\PYZus{}requirement}\PY{p}{(}\PY{l+s+s2}{\PYZdq{}x153\PYZdq{}}\PY{p}{,}\PY{l+s+s2}{\PYZdq{}variant\PYZdq{}}\PY{p}{,}\PY{l+s+s2}{\PYZdq{}example\PYZdq{}}\PY{p}{,}\PY{l+s+s2}{\PYZdq{}bzip\PYZdq{}}\PY{p}{,}\PY{l+s+s2}{\PYZdq{}True\PYZdq{}}\PY{p}{)}\PY{p}{.}
\PY{n+nf}{imposed\PYZus{}constraint}\PY{p}{(}\PY{l+s+s2}{\PYZdq{}x153\PYZdq{}}\PY{p}{,}\PY{l+s+s2}{\PYZdq{}depends\PYZus{}on\PYZdq{}}\PY{p}{,}\PY{l+s+s2}{\PYZdq{}example\PYZdq{}}\PY{p}{,}\PY{l+s+s2}{\PYZdq{}bzip2\PYZdq{}}\PY{p}{)}
\end{Verbatim}


\vspace{0.5em}
The \texttt{condition\_requirement} follows a similar pattern to \texttt{attr}, in
that it is a higher-order, variable arity relation to allow for uniform handling
of the various requirements.

In order to attach attributes to nodes conditionally (\textit{i.e.,} when a
\texttt{when} spec is satisfied), the concretizer includes a set of rules for
conditionally imposed constraints on nodes. These rules hinge on two relations,
\texttt{impose} and \texttt{imposed\_constraint}. The latter is a variable-arity
relation taking advantage of the higher order encoding of \texttt{attr} indexed by
a unique identifier, while the former controls constraints associated with that
identifier that are imposed on a particular node. What follows is an example of how
these facts are used in the logic program for the concretizer:

\begin{Verbatim}[commandchars=\\\{\},fontsize=\footnotesize,frame=single]
\PY{n+nf}{attr}\PY{p}{(}\PY{n+nv}{Attr}\PY{p}{,} \PY{n+nf}{node}\PY{p}{(}\PY{n+nv}{Name}\PY{p}{)}\PY{p}{,} \PY{n+nv}{Arg1}\PY{p}{)} \PY{p}{:\PYZhy{}}
  \PY{n+nf}{impose}\PY{p}{(}\PY{n+nv}{ID}\PY{p}{,} \PY{n+nf}{node}\PY{p}{(}\PY{n+nv}{Name}\PY{p}{)}\PY{p}{)}\PY{p}{,}
  \PY{n+nf}{imposed\PYZus{}constraint}\PY{p}{(}\PY{n+nv}{ID}\PY{p}{,} \PY{n+nv}{Name}\PY{p}{,} \PY{n+nv}{Arg}\PY{p}{)}\PY{p}{,} \PY{n+nf}{node}\PY{p}{(}\PY{n+nv}{Name}\PY{p}{)}\PY{p}{.}
\PY{n+nf}{attr}\PY{p}{(}\PY{n+nv}{Attr}\PY{p}{,} \PY{n+nf}{node}\PY{p}{(}\PY{n+nv}{Name}\PY{p}{)}\PY{p}{,} \PY{n+nv}{Arg1}\PY{p}{,} \PY{n+nv}{Arg2}\PY{p}{)} \PY{p}{:\PYZhy{}}
  \PY{n+nf}{impose}\PY{p}{(}\PY{n+nv}{ID}\PY{p}{,} \PY{n+nf}{node}\PY{p}{(}\PY{n+nv}{Name}\PY{p}{)}\PY{p}{)}\PY{p}{,}
  \PY{n+nf}{imposed\PYZus{}constraint}\PY{p}{(}\PY{n+nv}{ID}\PY{p}{,} \PY{n+nv}{Name}\PY{p}{,} \PY{n+nv}{Arg1}\PY{p}{,} \PY{n+nv}{Arg2}\PY{p}{)}\PY{p}{,} \PY{n+nf}{node}\PY{p}{(}\PY{n+nv}{Name}\PY{p}{)}\PY{p}{.}
\end{Verbatim}


\vspace{0.5em}
Note that there are similar rules at every potential arity of the \texttt{attr}
relation. \texttt{imposed\_constraint} is used to encode various package facts,
especially for reusable specs and \texttt{impose} is derived when the conditions
associated with the \texttt{ID} are present. In particular to this work, the
\texttt{ID} considered will be that of the hash of a reusable spec.

\subsubsection{Reusing concrete specs}
\label{sec:concretizer-reuse}
In order to reuse concrete specs, the concretizer must be aware of installed specs
indexed by their unique hash) and which constraints would
be imposed in their reuse. This is encoded directly by providing an
\texttt{installed\_hash} relation that ties package names to the hashes of
concrete specs and \texttt{imposed\_constraint} facts for all of the attributes
of the concrete spec. What follows is the encoding of the concrete spec from
section~\ref{sec:spack-concretizer}:

\begin{Verbatim}[commandchars=\\\{\},fontsize=\footnotesize,frame=single]
\PY{n+nf}{installed\PYZus{}hash}\PY{p}{(}\PY{l+s+s2}{\PYZdq{}example\PYZdq{}}\PY{p}{,} \PY{l+s+s2}{\PYZdq{}abcdef1234\PYZdq{}}\PY{p}{)}\PY{p}{.}
\PY{n+nf}{imposed\PYZus{}constraint}\PY{p}{(}\PY{l+s+s2}{\PYZdq{}abcd1234\PYZdq{}}\PY{p}{,}\PY{l+s+s2}{\PYZdq{}version\PYZdq{}}\PY{p}{,}\PY{l+s+s2}{\PYZdq{}example\PYZdq{}}\PY{p}{,}\PY{l+s+s2}{\PYZdq{}1.1.0\PYZdq{}}\PY{p}{)}\PY{p}{.}
\PY{n+nf}{imposed\PYZus{}constraint}\PY{p}{(}\PY{l+s+s2}{\PYZdq{}abcd1234\PYZdq{}}\PY{p}{,}\PY{l+s+s2}{\PYZdq{}variant\PYZdq{}}\PY{p}{,}\PY{l+s+s2}{\PYZdq{}example\PYZdq{}}\PY{p}{,}\PY{l+s+s2}{\PYZdq{}bzip\PYZdq{}}\PY{p}{,}\PY{l+s+s2}{\PYZdq{}True\PYZdq{}}\PY{p}{)}\PY{p}{.}
\PY{n+nf}{imposed\PYZus{}constraint}\PY{p}{(}\PY{l+s+s2}{\PYZdq{}abcd1234\PYZdq{}}\PY{p}{,}\PY{l+s+s2}{\PYZdq{}node\PYZus{}os\PYZdq{}}\PY{p}{,}\PY{l+s+s2}{\PYZdq{}example\PYZdq{}}\PY{p}{,}\PY{l+s+s2}{\PYZdq{}centos8\PYZdq{}}\PY{p}{)}\PY{p}{.}
\PY{n+nf}{imposed\PYZus{}constraint}\PY{p}{(}\PY{l+s+s2}{\PYZdq{}abcd1234\PYZdq{}}\PY{p}{,}\PY{l+s+s2}{\PYZdq{}node\PYZus{}target\PYZdq{}}\PY{p}{,}\PY{l+s+s2}{\PYZdq{}example\PYZdq{}}\PY{p}{,}\PY{l+s+s2}{\PYZdq{}skylake\PYZdq{}}\PY{p}{)}\PY{p}{.}
\PY{n+nf}{imposed\PYZus{}constraint}\PY{p}{(}\PY{l+s+s2}{\PYZdq{}abcd1234\PYZdq{}}\PY{p}{,}\PY{l+s+s2}{\PYZdq{}depends\PYZus{}on\PYZdq{}}\PY{p}{,}\PY{l+s+s2}{\PYZdq{}example\PYZdq{}}\PY{p}{,}\PY{l+s+s2}{\PYZdq{}bzip2\PYZdq{}}\PY{p}{)}\PY{p}{.}
\PY{n+nf}{imposed\PYZus{}constraint}\PY{p}{(}\PY{l+s+s2}{\PYZdq{}abcd1234\PYZdq{}}\PY{p}{,}\PY{l+s+s2}{\PYZdq{}hash\PYZdq{}}\PY{p}{,}\PY{l+s+s2}{\PYZdq{}bzip2\PYZdq{}}\PY{p}{,}\PY{l+s+s2}{\PYZdq{}1234abcd\PYZdq{}}\PY{p}{)}\PY{p}{.}
\end{Verbatim}


\vspace{0.5em}
We will update this encoding in section~\ref{sec:impl-new-encoding} with some
indirection to better facilitate splicing.

In order to select reusable specs, the concretizer may select at most one spec
for every node with a potential candidate (\textit{i.e.,} the \texttt{installed\_hash} fact
is present for a node with that name). If chosen, the concretizer imposes all
constraints associated with that concrete spec. Any node which is not selected
for reuse is then marked to be built instead. Then, the highest optimization
objective in Spack's concretizer is to reuse as many specs as possible, or
equivalently to minimize the number of specs needing to be built:
\par\vspace{0.5em}\noindent


\begin{Verbatim}[commandchars=\\\{\},fontsize=\footnotesize,frame=single]
\PY{p}{\PYZob{}} \PY{n+nf}{attr}\PY{p}{(}\PY{l+s+s2}{\PYZdq{}hash\PYZdq{}}\PY{p}{,} \PY{n+nf}{node}\PY{p}{(}\PY{n+nv}{Name}\PY{p}{)}\PY{p}{,} \PY{n+nv}{Hash}\PY{p}{)}\PY{l+s+sAtom}{:} \PY{n+nf}{installed\PYZus{}hash}\PY{p}{(}\PY{n+nv}{Name}\PY{p}{,} \PY{n+nv}{Hash}\PY{p}{)} \PY{p}{\PYZcb{}} \PY{l+m+mi}{1}
  \PY{p}{:\PYZhy{}} \PY{n+nf}{node}\PY{p}{(}\PY{n+nv}{Name}\PY{p}{)}\PY{p}{.}
\PY{n+nf}{impose}\PY{p}{(}\PY{n+nv}{Hash}\PY{p}{,} \PY{n+nv}{Node}\PY{p}{)} \PY{p}{:\PYZhy{}} \PY{n+nf}{attr}\PY{p}{(}\PY{l+s+s2}{\PYZdq{}hash\PYZdq{}}\PY{p}{,} \PY{n+nv}{Node}\PY{p}{,} \PY{n+nv}{Hash}\PY{p}{)}\PY{p}{,} \PY{n+nv}{Node}\PY{p}{.}
\PY{n+nf}{build}\PY{p}{(}\PY{n+nv}{Node}\PY{p}{)} \PY{p}{:\PYZhy{}} \PY{n+nv}{Node}\PY{p}{,} \PY{o}{not} \PY{n+nf}{attr}\PY{p}{(}\PY{l+s+s2}{\PYZdq{}hash\PYZdq{}}\PY{p}{,} \PY{n+nv}{Node}\PY{p}{,} \PY{n+nv}{Hash}\PY{p}{)}\PY{p}{.}
\PY{l+s+sAtom}{\PYZsh{}}\PY{l+s+sAtom}{minimize} \PY{p}{\PYZob{}} \PY{l+m+mi}{100}\PY{p}{,} \PY{n+nv}{PackageNode} \PY{l+s+sAtom}{:} \PY{n+nf}{build}\PY{p}{(}\PY{n+nv}{PackageNode}\PY{p}{)} \PY{p}{\PYZcb{}}
\end{Verbatim}


\vspace{0.2em}
\subsection{Augmenting Spack's Package Language}
\label{sec:impl-new-directive}
\subsubsection{Representation in the packaging DSL}
We add a single directive to Spack's packaging language for specifying
ABI-compatibility relationships between packages, \texttt{can\_splice}.  The
bottom of fig.~\ref{fig:example-pkg} demonstrates \texttt{can\_splice} directives
for our example package. The first spec argument describes the constraints on
the target of the splice (\textit{i.e.,} the spec that would be replaced by the splice) and
the \texttt{when} spec argument describes the constraints on the package for the
splice to be valid. Both arguments support the full spec syntax. Note that
splices can be specified between two versions of the same package as well as
different packages. \texttt{can\_splice} is a natural extension to the
packaging language since it utilizes the same API as most other Spack
directives.

This directive inverts the dependency structure of Spack when
considering binary distributions of packages. While the \texttt{depends\_on}
relationship is given by the \emph{depending} package, \texttt{can\_splice} allows
\emph{dependent} packages to state when it would not be ABI-breaking for them to
be swapped for another package. One subtlety to note is that packages state which
specs they \emph{can replace} rather than by which they \emph{can be replaced}.
This is an acknowledgment of the fact that developers of a common ABI, like
that of \textsc{MPICH}, will not necessarily be aware of every ABI-compatible
replacement, but developers of these replacements will know that their package
is ABI compatible.

\subsubsection{Encoding in ASP}
Unlike other directives in Spack, rather than being encoded as facts to describe
the constraints imposed by the \texttt{can\_splice} directive, we choose to
encode the constraints from \texttt{can\_splice} as a specialized rule.
Figure~\ref{fig:can-splice-encoding} demonstrates our ASP encoding for the second
\texttt{can\_splice} directive in fig.~\ref{fig:example-pkg}. It is a single rule
that derives the fact \texttt{can\_splice}. This fact represents that the package
node in its first argument can replace a reusable spec with the name of the
second argument and the hash of the third argument, and can only be derived if
there is a package node satisfying the \texttt{when} spec and a pre-built spec
satisfying the target. It is a near-direct translation of the
\texttt{can\_splice} directive to ASP. Being able to match node attributes to
pre-built spec attributes is one of factors that motivated the change in
encoding of reusable specs to use \texttt{hash\_attr}.

\subsection{Changing the Encoding of Reusable Specs}
\label{sec:impl-new-encoding}
Recall from section~\ref{sec:concretizer-reuse} that reusable specs were encoded by
directly as \texttt{imposed\_constraint} facts for every attribute of the concrete
spec, indexed by its hash. Since splicing changes the dependencies of a concrete
spec, the concretizer needs to be able to change which dependencies are imposed
by the spliced spec. Yet, in the previous encoding, all constraints from the
original spec are imposed immediately, and ASP does not have a mechanism for
removing derived facts from a model. Therefore, we changed the encoding of
pre-installed specs to add a layer of indirection between the encoding of the
attributes of reusable specs, and when those attributes are actually imposed.
%


\begin{figure}
  \begin{subfigure}[t]{1.0\columnwidth}

\input{code8.tex}
\caption{The new encoding of reusable specs.}
    \label{fig:new-encoding-hash-attrs}
  \end{subfigure}
  \begin{subfigure}[b]{1.0\columnwidth}

\input{code9.tex}
\caption{Rules that recover the \texttt{imposed\_constraint} facts from the original
encoding of reusable specs.}
    \label{fig:new-encoding-imposed-constraints}
  \end{subfigure}
  \caption{The new encoding of reusable specs introduces indirection into the of \texttt{imposed\_constraint}, this indirection is key for splicing to allow for
  splicing to change the dependencies of reusable specs.}
\label{fig:new-encoding}
\Description{ASP code demonstrating the change in encoding to already installed specs.}
\end{figure}


%
Figure~\ref{fig:new-encoding-hash-attrs} demonstrates how we introduce this
indirection through \texttt{hash\_attr} facts and
fig.~\ref{fig:new-encoding-imposed-constraints} shows how we recover the semantics
from the previous encoding. Note that the initial imposition of \texttt{"hash"} and
\texttt{"depends\_on"} depends on the presence of
\texttt{can\_splice}. These attributes are where splices can be
introduced and this encoding allows the concretizer to make decisions about
imposition and reuse of either the original dependency or a spliced dependency.
Without this \texttt{can\_splice}, there are no valid splices and
thus the concretizer defaults to its previous behavior of always reusing the
dependencies of any pre-built spec whenever it reuses the spec itself.
\subsection{Splicing in the Concretizer}
\label{sec:impl-splice-concretizer}


\begin{figure}
\begin{subfigure}[t]{1.0\columnwidth}

\input{code11.tex}
\caption{Compiled ASP encoding of the \texttt{can\_splice} directive included
  as a comment on the first line of the fragment}
\label{fig:can-splice-encoding}
\end{subfigure}
\begin{subfigure}[b]{1.0\columnwidth}

\input{code12.tex}
\caption{The core logic implementing splicing. This fragment selects whether to
  execute a splice based on the presence of the \texttt{can\_splice} fact, and then
  imposes the new spliced dependency instead of the pre-installed spec's original
  dependency.} 
  \label{fig:splice-lp}
\end{subfigure}
\caption{The splicing implementation in ASP portion of Spack's concretizer.}
\label{fig:splice-concretizer}
\Description{An ASP snippet implementing the core of the splicing logic.}
\end{figure}


%
Figure~\ref{fig:splice-lp} demonstrates the core fragment of the concretizer
implementing splicing. The rules can be read top to bottom as implementing the
fairly simple decision procedure for splicing. The rules beginning on lines 1
and 4 describe an exclusive-or condition for how to impose the dependencies of the spec
that is being chosen for reuse provided it has a potential candidate for a
splice. When reusing a concrete spec, the concretizer may impose its original
dependency (the choice on line 1). If it does not (the condition on line 8), it
must instead splice in a compatible dependency (the conclusion on line 4). If it
decides to splice in a new dependency (the condition on line 12), that new
dependency is imposed instead of the original one (the conclusion on line 9).

Perhaps it is surprising how little logic is required to implement the core
splicing in the concretizer's logic program. We owe this elegance to the change
in encoding of reusable concrete specs. The small indirection added through
\texttt{hash\_attr} facts provides the perfect hook for changing the dependency
when splicing. By refactoring machinery already present in the concretizer (\textit{i.e.,}
\texttt{impose} and \texttt{imposed\_constraint}), we are still able to take advantage
of the logic that maintains the integrity of the whole spec-DAG (\textit{e.g.,} ensuring
compatible microarchitectures among all specs). After making this encoding
change, Spack core developers have found subsequent uses in the concretizer for
indirection in imposing concrete specs\footnote{https://github.com/spack/spack/pull/45189}.

Ultimately, the concretizer produces a mapping between pre-built specs (\textit{i.e.,}
nodes with the \texttt{"hash"} attribute) and their dependencies that will be
replaced by splicing (\textit{i.e.,} nodes with the \texttt{"splice"} attribute). This
mapping can be used directly not only to build the spliced specs in a
straightforward manner but also in the rewiring of spliced specs, as
described in section~\ref{sec:splice-relocation}. Finally, the splicing logic can be
conditionally loaded, and thus is a fully transparent opt-in feature. We
validate that our encoding change for pre-installed specs incurs minimal
overhead in the next section.



\section{Performance Results}

  \begin{figure*}[!t]
    \includegraphics[trim=0 500 0 0, clip]{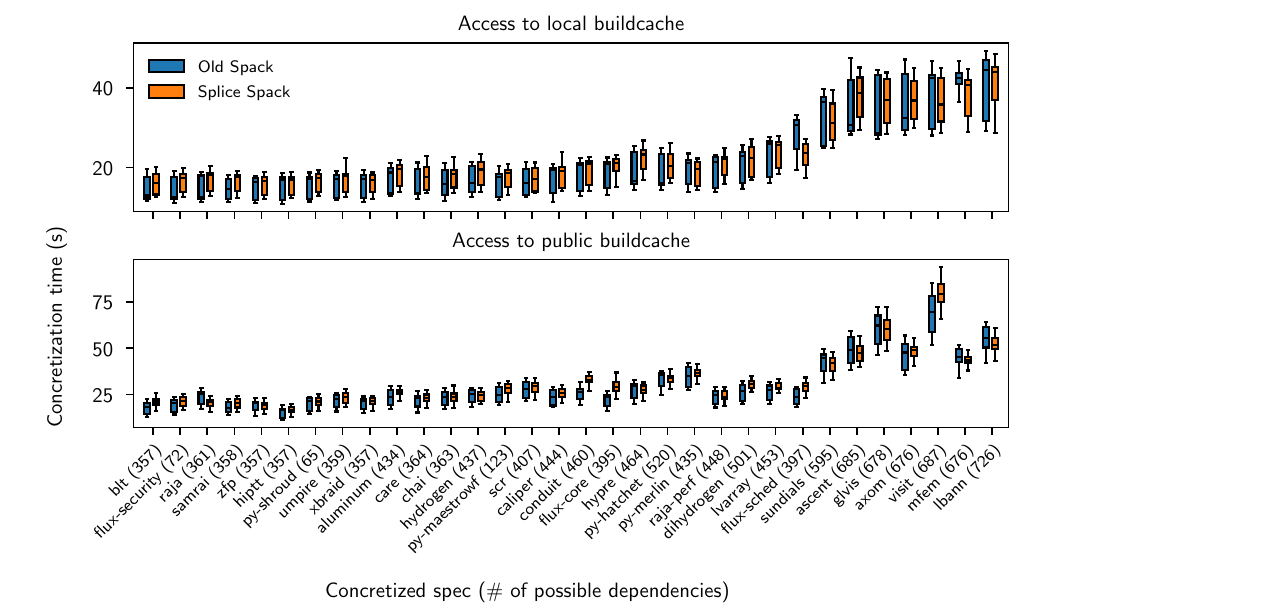}
    \caption{A comparison of the performance of \textit{old spack} and \textit{splice spack} with
      respect to their encoding of reusable specs over 30 runs.}
    \label{fig:encoding-perf}
    \Description{A box and whisker plot demonstrating the impact of the encoding change on concretizer performance.}
  \end{figure*}%



  \begin{figure*}[!t]
    \includegraphics[trim=0 500 0 0, clip]{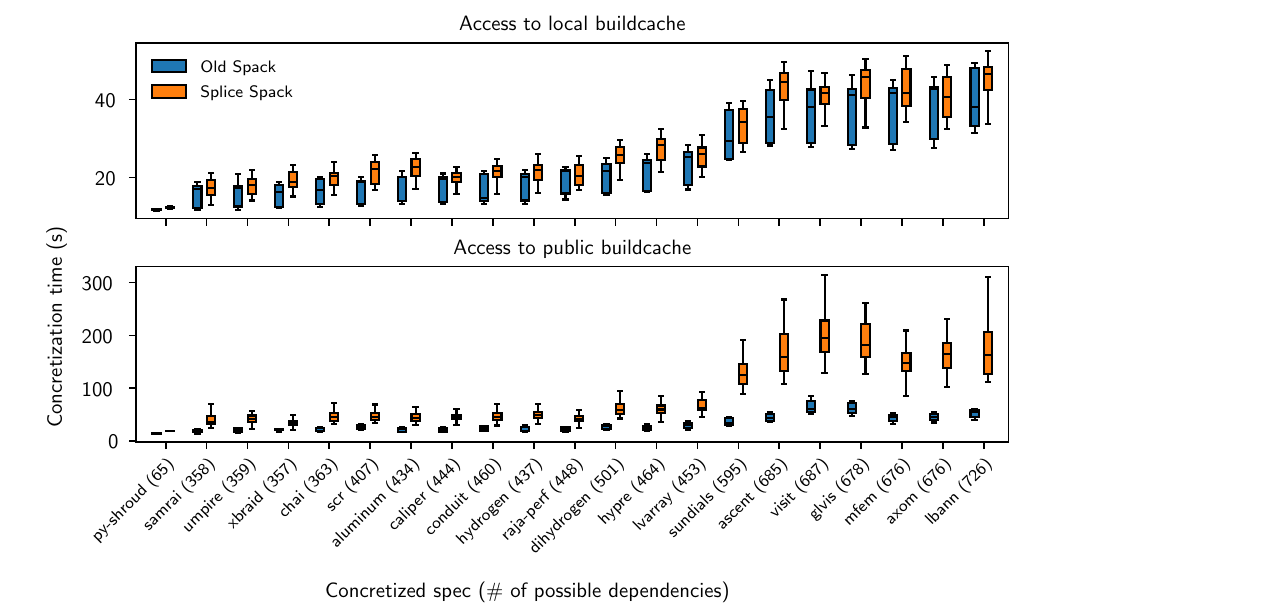}
    \caption{A comparison of the performance of a non-spliced solution for \textit{old spack} with \textit{splice spack} considering spliced solutions over 30 runs.}
    \label{fig:splice-perf}
    \Description{A box and whisker plot demonstrating the overhead introduced by considering splices.}
  \end{figure*}%



  \begin{figure*}[!t]
    \includegraphics[trim=0 600 0 0, clip]{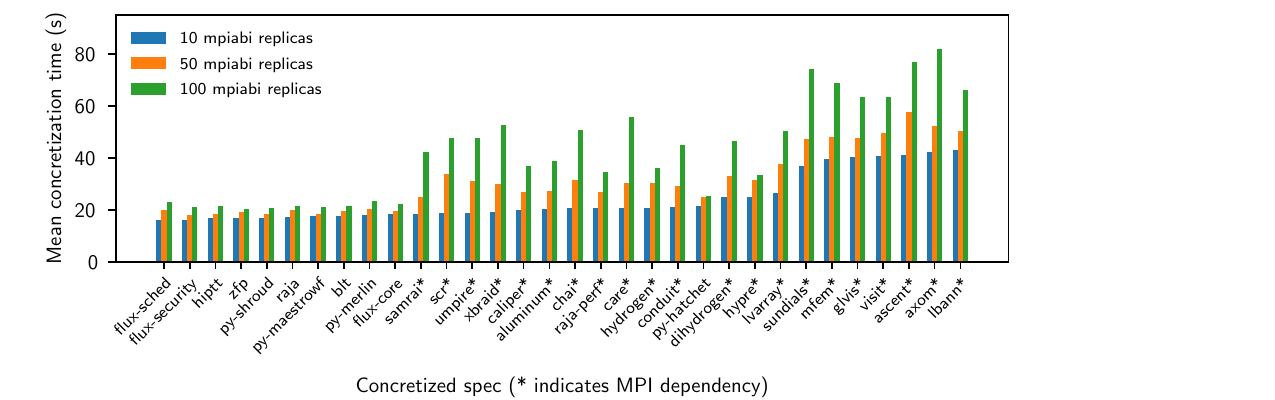}
    \caption{An evaluation of the scaling behavior of \textit{splice spack} when the
      number of potential splice candidates is increased.}
    \label{fig:splice-scaling}
    \Description{A bar chart demonstrating the overhead of introducing a number of splice candidates.}
  \end{figure*}%

We have fully implemented automatic splicing in Spack's concretizer and aim to
answer 4 questions about its performance:
\begin{enumerate}
\item [RQ1.]Do the changes to the encoding of prebuilt packages in the concretizer
  introduce bugs or performance regressions? \label{rq:dev-regressions}
\item [RQ2.] Does the concretizer produce spliced solutions when
  necessary? \label{rq:splice-correctness}
\item [RQ3.] What is the overhead of considering spliced
  solutions? \label{rq:splice-performance}
\item [RQ4.] How does automatic splicing performance scale in relation to the number of
  candidate splices? \label{rq:splice-scaling}
\end{enumerate}

\subsection{Experimental Setup}
\subsubsection{System Setup}
All performances runs were executed on a single node with 96 Intel Xenon Gold
6342 (Icelake) processors and 1TB of memory running Ubuntu 22.04.

\subsubsection{The RADIUSS stack and the mpiabi mock package}
In each of our experiments we consider the concretization of specs from the
RADIUSS software stack. RADIUSS is an open-source collection of HPC
infrastructure (\textit{e.g.,} \textsc{Flux} and \textsc{LvArray}), portability tools
(\textit{e.g.,} \textsc{RAJA} and \textsc{CHAI}), data management and visualization
(\textit{e.g.,} \textsc{glvis} and \textsc{hatchet}), and simulation applications
(\textit{e.g.,} \textsc{ascent} and \textsc{sundials}) created by LLNL in order to
provide a unified foundation for developing HPC applications.  Concretizing
RADIUSS packages is a common use-case for Spack and the packages vary in both
overall dependency structure and number of dependencies. Furthermore, many of
the packages in the RADIUSS stack have a virtual dependency on MPI which serves
as our splice target. For a splice candidate, we created a mock package,
\textsc{mpiabi}, based on \textsc{mvapich}, with a single version and the
ability to splice into \texttt{mpich@3.4.3}. In section~\ref{sec:scaling}, we introduce
many copies of this package in order to assess the scaling factor of the number
of splice candidates.

\subsubsection{The local and public buildcaches}
Splicing requires the existence of pre-concretized specs so we use
\textit{buildcaches} of specs.  Buildcaches are Spack's way of providing
reusable binaries of prebuilt specs that map concrete specs to their build
targets (\textit{e.g.,} shared libraries, executables, \textit{etc.}) and whose contents can be
reasoned about by the concretizer when deciding which specs can be reused. Our
experiments rely on a local and a public buildcache. The local buildcache
consists of only specs from the RADIUSS stack (and transitive dependencies),
allowing us to construct a controlled environment to evaluate the correctness of
our implementation, containing \textasciitilde{}200 specs. The public buildcache, which is
generated by Spack \footnote{www.cache.spack.io} as a continuous integration
task, contains over 20,000 prebuilt specs that include multiple configurations
of the RADIUSS stack, which provides a use-case more in line with that of
the average Spack user.

\subsubsection{Concretization Objectives and Configurations}
We consider the concretization of each of the 32 individual specs in the RADIUSS
stack and there are three axes for configuration in our experiments. The first
axis is the version of Spack under consideration. We compare our implementation,
hence \textit{splice spack}, to the last commit on the main Spack Github repository
prior to the introduction of automatic splicing, hence
\textit{old spack}. \footnote{Git commit hash for \textit{old spack}:
  ad518d975c711c04bdc013363d8fc33a212e9194, git commit hash for \textit{splice spack}:
  9f7cff1780d1e3e97cf957d686966a74d3840af6. Spack code is available at
  \url{https://github.com/spack/spack}}. The second axis is the number of
reusable specs the concretizer considers when producing solutions. This can be
either the local RADIUSS buildcache or the public buildcache. The third axis is
whether automatic splicing is enabled when testing that implementation. Recall
from section~\ref{sec:implementation} that \textit{splice spack} allows for users to
conditionally enable the concretizer to consider automatic splices; however,
disabling automatic splicing \textit{does not} revert to the previous encoding
of reusable specs. Therefore, experiments where the feature is disabled are to
evaluate the performance of the new encoding of reusable specs without the added
expressivity of allowing splicing. Enabling splicing allows the concretizer to
consider solutions that involve splicing, but does not explicitly constrain it
to only consider solutions with splicing. In order to address the variability
inherent in heuristic nature of Spack's dependency resolution engine, we
evaluate each concretization objective in each configuration 30 times.

\subsection{Determining the Impact of the Changed Encoding for Reusable Specs}
\label{sec:encoding-exp}
In order to answer \ref{rq:dev-regressions}, we consider whether our
changes to the encoding of reusable specs and strategy for reuse
introduce regressions into the concretizer.  To accomplish this, we concretize
each objective using both \textit{old spack} and \textit{splice spack} against both
buildcache configurations. We consider only the case where the automatic
splicing feature is disabled, such that the only difference between the two
implementations is the encoding of reusable specs.
\par
Figure~\ref{fig:encoding-perf} shows the distribution over 30 runs
for concretizing the entire RADIUSS stack in the four possible
configurations. Across all specs in the RADIUSS stack, we see an
4.7 percent increase in average concretization time with access
to the local buildcache from our encoding change, and a 7.1
percent increase with access to the public buildcache. Therefore, we can see
that the change to the encoding of resuseable specs only introduces neglible
overhead into the performance of the concretizer.

\subsection{Evaluating the Correctness and Performance of Automatic Splicing}
\label{sec:corr-perf-exps}
In order to answer \ref{rq:splice-correctness} and \ref{rq:splice-performance},
we consider the subset of specs in RADIUSS that have a virtual dependency on
MPI. We concretize these MPI-dependent specs separately and jointly with an
explicit dependency on either \texttt{mpich} in the case of \textit{old spack}, or our
mock MPI package (\texttt{mpiabi}) in the case of \textit{splice spack} using both
buildcache configurations. We ensure that \textit{splice spack} produces spliced
solutions when possible. We also include \texttt{py-shroud} in order to evaluate
the overhead of enabling splicing on specs which cannot be spliced. Note that
\textit{splice spack} is solving a harder problem than \textit{old spack}, since it now must
consider splicing rather than simple reuse, and thus one should expect a
increase in concretization time.

Figure~\ref{fig:splice-perf} shows the distribution over 30 runs for
concretizing \texttt{py-shroud} and the MPI-dependent specs in the RADIUSS stack
in the four possible configurations. Across all MPI-dependent specs in the
RADIUSS stack, we see a 17.1 percent increase in
concretization time with access to the local buildcache from our encoding
change, and a 153 percent increase with access to the public
buildcache. We see virtually no difference in concretization time for
\texttt{py-shroud} from our extension. Recall that the public buildcache is
roughly 2 orders of magnitude larger than the local buildcace (~200 specs
vs. ~20,000 specs).

Thus, while enabling splicing introduces a potentially two minute increase to
concretization time in the case of specs like \texttt{visit} and \texttt{glvis}
against the large public buildcache, every spliced solution could save potential
hours of time spent building software. Furthermore, since splicing can be
conditionally enabled, users can easily opt out if they would not benefit from
splicing and would prefer faster concretization.
 
\subsection{Scaling the Number of Spliceable Packages}
\label{sec:scaling}
The previous experiments consider splicing a single package
with a single \texttt{can\_splice} directive targeting \texttt{mpich@3.4.3};
however, in more realistic scenarios there may be many potential splice
candidates for a given target. Therefore, in order to answer
\ref{rq:splice-scaling}, we will consider the same concretization goals as
section~\ref{sec:corr-perf-exps}, but instead of varying the version of Spack, we will
instead scale the number of potential splice candidates. We accomplish this by
first creating 100 of copies of \texttt{mpiabi} differing only in name. We then
concretize with the same objectives as the previous experiment using only the
local buildcache while giving the concretizer access to increasingly large
subsets of the 100 mock packages. We also require that concretized specs
\emph{do not} depend on \texttt{mpich}, but do not constrain which of the replicas
the concretizer chooses.

Figure~\ref{fig:splice-scaling} shows how the average concretization time scales for
concretization of the RADIUSS stack. Across all MPI-dependent packages, the
average percent increase in concretization time between 10 and 100 replicas is
74.2. Again, this increase on the order of minutes is
inconsequential compared the potential hours saved by not having to rebuild
packages. Note that there is very little overhead in scaling the number of
replicas for specs which do not have an MPI dependency.



\section{Related Work}
\label{sec:related}

Splicing generalizes approaches to avoid rebuilds in source-based package management by
allowing binary substitution across ABI-compatible implementations. Related systems
either avoid this problem through rigid policies or handle it with limited,
use-case-specific tooling. Systems like {\tt rpm}, {\tt apt}, {\tt yum}, {\tt dnf}, and
other binary package managers, regularly resolve dependencies without rebuilding their
dependents. Typical Linux distributions upgrade packages in-place, and when a dependency
binary is installed, its dependents are rebuilt. Traditional distributions {\it trust}
the maintainers to ensure that ABI compatibility is maintained. Some, like Red Hat, use
tooling like {\tt libabigail} to ensure that new packages adhere to the ABI of existing
packages. Fundamentally, all of these solutions rely on the maintenance process to
ensure ABI compatibility. Moreover, these systems do not reason deeply about ABI--they
typically only manage version constraints, and do not reason about flags, build options,
or microarchitectures as our conditional {\tt can\_splice} solution allows.

So-called functional package managers like Nix~\cite{dolstra2004nix,dolstra2008nixos}
and Guix~\cite{courtes-guix-2015} ensure that ABI compatibility is preserved using a
convention similar to Spack. They identify packages based on their own configuration
{\it and} that of all their dependencies, and they traditionally force full rebuilds if
any dependencies change. Nix mitigates this issue by providing a public
(non-relocatable) binary cache full of package substitutes. Again, maintainers ensure
that rebuilds happen and packagers can easily install once all rebuilds happen.

Guix has implemented a solution for the cascading rebuild problem most similar to our
own; their variant is called {\it grafts}~\cite{guix-manual-grafts}. Their solution is
intended for security updates and one-time ABI replacements.
Splicing lets package authors declare ABI compatibility using spec constraints, enabling
the solver to select compatible replacements based on context. Unlike grafts, these
declarations live with the replacing package, not the replaced one.
Guix allows developers to specify replacements for their own package that avoid cascading
rebuilds~\cite{courtes2016timely,courtes2020grafts}. This
solution differs from ours in that the {\it replaced} package specifies what can replace
it. This effectively means that the replaced package must {\it know} about possible
replacements, and that only one replacement can be specified at any given time. This
makes sense for Guix's use case of security updates. If a CVE is discovered, a grafted
package is replaced with a ``fixed'' version. However, this implementation is unsuitable
for implementations like MPI, where there may be {\it many} possible ABI-compatible
replacements that may be different depending on the host platform and the choices of the
developer maintaining the software stack. We could not express in Guix, for example,
that \textsc{Cray MPICH}, \textsc{mvapich}, and \textsc{intel-mpi} are all suitable
replacements for \textsc{mpich}. Further, Guix has no solver, so there is no reasoning
on the part of the package manager about ABI constraints -- the maintainers must ensure
that any specified replacement is ABI-compatible with the replaced package.
Splicing instead allows for
an ecosystem in which new replacements can be easily added, users can choose which
replacement they use, and the solver ultimately ensures the safety of user choices.



\section{Conclusions and Future Work}
There is a fundamental tension between source-based and binary-based package
management trading off configurability for installation speed. This tension
arises from a lack of explicitly modeling ABI-compatibility in source-based
package managers, while concerns of ABI-compatibility limit configurability in
binary-based package managers.

In this paper, we provide an extension to the source-based Spack package manager
that allows for explicitly stating and reasoning about ABI-compatibility among
packages and automatically generating solutions with a minimal number of builds
by reusing and relinking ABI-compatible packages when possible. Our extension
incurs minimal overhead, while greatly extending the modeling capability of
Spack. We have also demonstrated that our approach scales well in the presence
of many opportunities for splicing. Our hope is that this extension will increase
the efficiency of Spack users, and provide a testing bed for future research on
ABI compatibility.

Currently, ABI compatibility must be specified by package
developers manually adding \texttt{can\_splice} to their package classes. The
one-time effort of modeling a package's ABI can provide massive efficiency gains
for all downstream users of a package. In the future, we will develop methods
for automating ABI discovery for the Spack ecosystem in order to reduce
developer burden.



\begin{acks}
  This material is based upon work supported by the U.S. Department of Energy,
  Office of Science under Award Number DESC0025613 and Lawrence Livermore National
  Laboratory under contract DE-AC52-07NA27344.

  \emph{Disclaimer}: This report was prepared as an account of work sponsored by
  an agency of the United States Government. Neither the United States
  Government nor any agency thereof, nor any of their employees, makes any
  warranty, express or implied, or assumes any legal liability or responsibility
  for the accuracy, completeness, or usefulness of any information, apparatus,
  product, or process disclosed, or represents that its use would not infringe
  privately owned rights. Reference herein to any specific commercial product,
  process, or service by trade name, trademark, manufacturer, or otherwise does
  not necessarily constitute or imply its endorsement, recommendation, or
  favoring by the United States Government or any agency thereof. The views and
  opinions of authors expressed herein do not necessarily state or reflect those
  of the United States Government or any agency thereof.
\end{acks}


\bibliographystyle{ACM-Reference-Format}




\end{document}